\documentclass[aps,prb,twocolumn,showpacs,amsmath,amssymb,superscriptaddress]{revtex4-1}

\usepackage{bm}
\usepackage{graphicx}
\usepackage[usenames,dvipsnames,svgnames,table]{xcolor}
\usepackage[colorlinks=true,linkcolor=RoyalBlue,citecolor=RoyalBlue]{hyperref}
\usepackage{textcomp}
\usepackage{soul}
\usepackage{pbox}

\newcommand{\bracket}[1]{\langle #1 \rangle}

\begin{document}

\title{Magnetic ground state of semiconducting transition metal trichalcogenide monolayers}

\author{Nikhil Sivadas}
\email{nsivadas@andrew.cmu.edu}
\affiliation{Department of Physics, Carnegie Mellon University, Pittsburgh, Pennsylvania 15213, USA}

\author{Matthew W. Daniels}
\affiliation{Department of Physics, Carnegie Mellon University, Pittsburgh, Pennsylvania 15213, USA}

\author{Robert H. Swendsen}
\affiliation{Department of Physics, Carnegie Mellon University, Pittsburgh, Pennsylvania 15213, USA}

\author{Satoshi Okamoto}
\affiliation{Materials Science and Technology Division, Oak Ridge National Laboratory, Oak Ridge, Tennessee 37831, USA}

\author{Di Xiao}
\email{dixiao@cmu.edu}
\affiliation{Department of Physics, Carnegie Mellon University, Pittsburgh, Pennsylvania 15213, USA}
\date{\today}

\begin{abstract}
Layered transition metal trichalcogenides with the chemical formula $ABX_3$ have attracted recent interest as potential candidates for two-dimensional magnets.  
Using first-principles calculations within density functional theory, we investigate the magnetic ground states of monolayers of Mn- and Cr-based semiconducting trichalcogenides.  
We show that the second and third nearest-neighbor exchange interactions ($J_2$ and $J_3$) between magnetic ions, which have been largely overlooked in previous theoretical studies, 
are crucial in determining the magnetic ground state.  
Specifically, we find that monolayer CrSiTe$_3$ is an antiferromagnet with a zigzag spin texture due to significant contribution from $J_3$, 
whereas CrGeTe$_3$ is a ferromagnet with a Curie temperature of 106 K. 
Monolayers of Mn-compounds (MnPS$_3$ and MnPSe$_3$) always show antiferromagnetic N{\'e}el order.  
We identify the physical origin of various exchange interactions, and demonstrate that strain can be an effective knob for tuning the magnetic properties. Possible magnetic ordering in the bulk is also discussed. Our study suggests that $ABX_3$ can be a promising platform to explore 2D magnetic phenomena.
\end{abstract}
\maketitle

\section{Introduction}

Recent years have seen a surge of interest in two-dimensional (2D) atomic crystals due to their highly tunable physical properties and immense potential in scalable device applications.~\cite{Novoselov04p666, Novoselov05p197, Zhang05p201,Geim07p183, Neto09p109}  Since the initial discovery of graphene, the family of 2D crystals has grown considerably with new additions such as boron nitride and transition metal dichalcogenides. The emergence of transition metal compounds in the landscape of 2D crystals is particularly advantageous as it opens the door to many physical properties not available in graphene.~\cite{Novoselov05p10451,Splendiani10p1271, PhysRevLett.105.136805,Radisavljevic11p147,Wang12p699,Geim13p419}  For example, in monolayer MoS$_2$ the large spin-orbit interaction leads to a unique spin-valley coupling which might be useful for spintronic applications.~\cite{Zhu11p153402,PhysRevLett.108.196802,PhysRevB.86.165108,Wenyu13p125301,Xu14p343}  Another interesting possibility brought by transition metal elements is magnetism,~\cite{McGuire15p612} a property still missing in the current line-up of 2D crystals.  In this regard, transition metal trichalcogenides (TMTC) such as MnPS$_3$ represents a rather attractive material family. Similar to dichalcogenides, these are layered compounds with weak interlayer Van der Waals interactions.  Furthermore, these materials are known to exhibit a large variety of magnetic phases,~\cite{Wildes98p6417,Wiedenmann81p1067,Brec86p3,PhysRevB.46.5425,Takano04E593,Siberchicot96p5863}   making them ideal candidates for exfoliated 2D magnets. The successful fabrication of a truly 2D magnet would also significantly advance our understanding of low-dimensional magnetism.

Despite the obvious interest and more than three decades of experimental studies of bulk TMTC, magnetism in these materials remains to be fully understood.  In particular, even though the spin wave measurement by inelastic neutron scattering has pointed out the importance of exchange interactions beyond nearest-neighbor (NN) spins,~\cite{Wildes98p6417,Wildes12p416004} relatively little is known about the nature of these interactions and their effect on the magnetic ground state.  Additionally, the 2D confinement of electrons upon exfoliation often leads to properties quite different from the bulk crystals.  It is thus interesting to ask whether there is any change of the magnetic ground state when these materials are thinned down to monolayers.

With these questions in mind, we investigate in this paper the magnetic ground states of monolayers of Mn- and Cr-based semiconducting TMTC, using first-principles calculations within the framework of density functional theory (DFT).  The Mn-compounds (MnPS$_3$ and MnPSe$_3$) are known to exhibit antiferromagnetic (AF) N{\'e}el order in their bulk form,~\cite{Wildes98p6417, Jeevanandam99p3563} and are chosen here as benchmark for our calculations due to the extensively available experimental data.  Interesting properties such as coupled spin and valley degrees of freedom have also been predicted for monolayers of these materials.~\cite{Li13p3738}
The Cr-compounds (CrSiTe$_3$ and CrGeTe$_3$), on the other hand, are reported to be ferromagnetic (FM) in bulk,~\cite{Carteaux95p251,Casto15p,Carteaux95p69} thus present a highly interesting system to realize 2D ferromagnets.  One of the motivations of the present paper is to provide some quantitative understanding of magnetism in these compounds. 

Our main findings are summarized below.  The majority of the paper is focused on monolayers.   We show that the second and third NN exchange interactions ($J_2$ and $J_3$) mediated through the $p$ states of chalcogen anions are crucial in determining the magnetic ground state.  Specifically, we find that monolayer CrSiTe$_3$ is an antiferromagnet with a zigzag spin texture due to significant contribution from $J_3$, whereas CrGeTe$_3$ a ferromagnet with a Curie temperature of 106 K.  This result is in sharp contrast with previous theoretical studies in which only the NN exchange interaction ($J_1$) was considered.~\cite{Xingxing14p7071,Chen15p60} We discuss the physical origin of various exchange interactions, and demonstrate that strain can be an effective knob for tuning the magnetic properties.  A uniform in-plane tensile strain of $\sim$ 3\% can tune the magnetic ground state of CrSiTe$_3$ from zigzag to ferromagnet, with a critical temperature of 111 K. We also find that in bulk CrSiTe$_3$, the intralayer magnetic ordering is very sensitive to the out-of-plane lattice constant. For the experimental out-of-plane lattice constant (21.0 {\AA}), the intralayer magnetic ordering is FM in nature. However, the interlayer coupling favors AF over FM coupling.  This is in contradiction with experimental results. ~\cite{Carteaux95p251,Casto15p}  One possible reason for this discrepancy  is discussed, but the actual mechanism for ferromagnetism in bulk CrSiTe$_3$ remains an open question.

\section{\label{sec:structure}Crystal and Magnetic Structure}

\begin{figure}
\includegraphics[width=1.0\columnwidth]{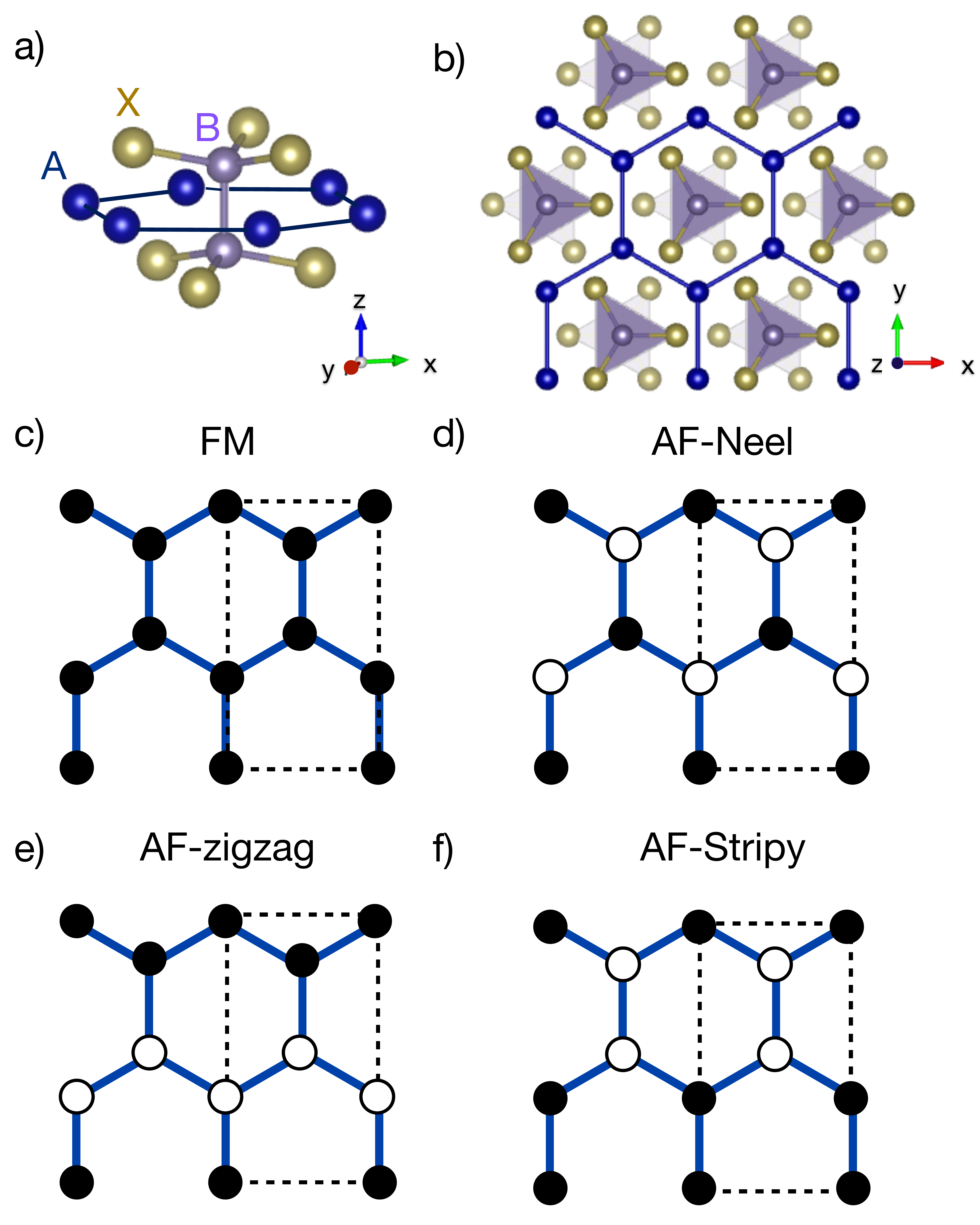}
\caption{\label{fig:Fig1} (Color online) Crystal and magnetic structure of transition metal trichalcogenides $ABX_3$. The crystal structure (a) and the top view (b) of monolayers of $ABX_3$.  
The transition metal $A$ atoms form a honeycomb structure with $B_2 X_6$ ligand occupying the interior of the honeycomb. 
Top view of the different spin configurations: the FM ordered (c), AF-N{\'e}el ordered (d), AF-zigzag ordered (e), AF-stripy ordered (f), with only the transition metal ions shown. 
Up (down) spins are represented by black filled-in (open) circles.  The crystal structure is drawn using VESTA.~\cite{Momma11p1272}}
\end{figure}

\begin{figure}
\includegraphics[width=\columnwidth]{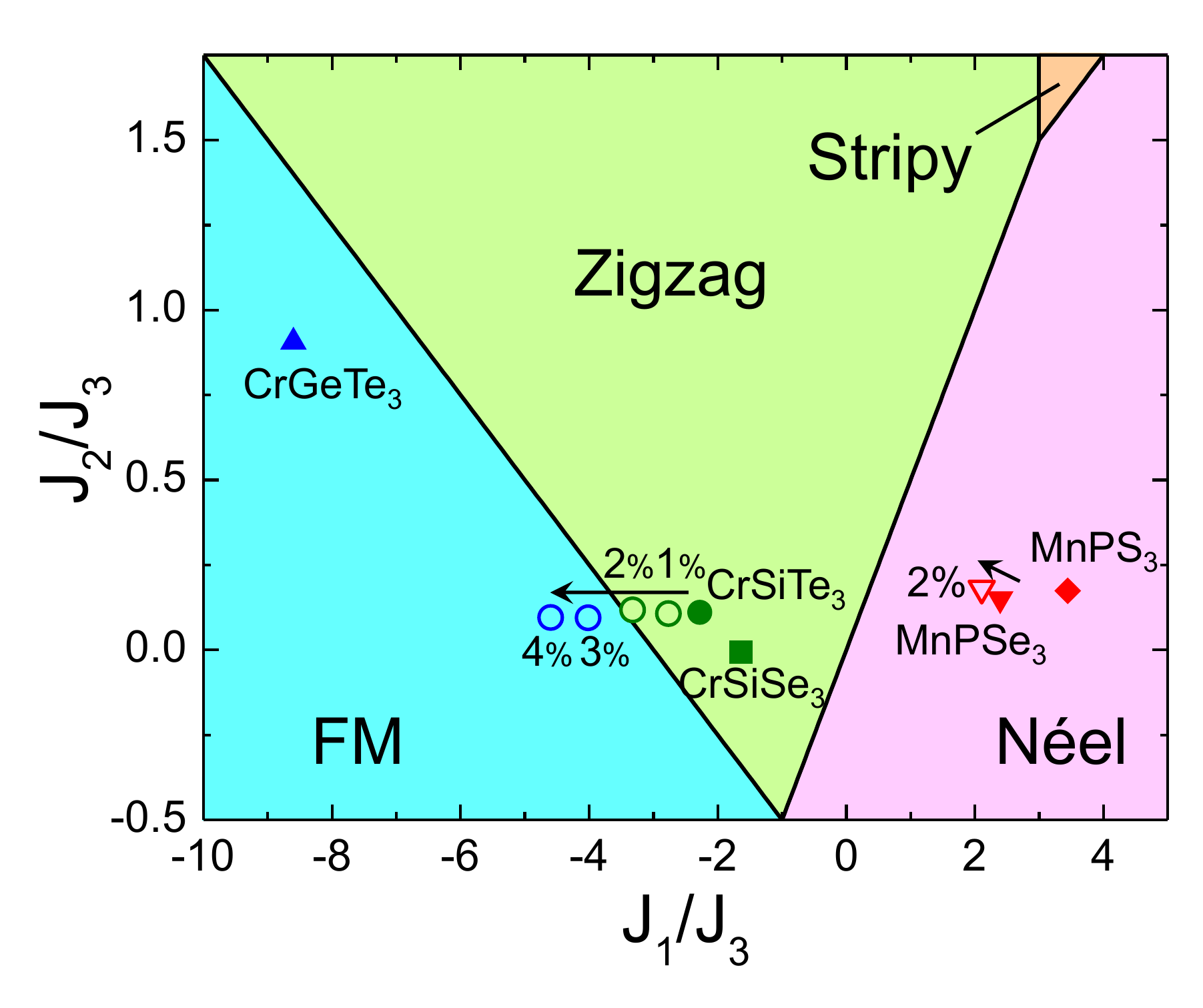}
\caption{\label{fig:J1_vs_J2} (Color online) 
 The ground state magnetic phase diagram for our spin model as a function of $J_1/J_3$ and $J_2/J_3$. 
Since our calculation finds $J_3$ to be always AF, only $J_3 > 0$ is considered. 
Spins are treated as classical degrees of freedom.
All compounds studied are located at corresponding parameter values.
Open symbols are positions under tensile strains with arrows indicating the change from the unstrained cases.}
\end{figure}

Transition metal trichalcogenides with the chemical formula $ABX_3$ are layered compounds with the structural space group of R$\overline{3}$, except MnPS$_3$, which forms monoclinic crystals with the $C2/m$ space group. In all compounds, the different layers are held together by weak van der Waals force.  It has been predicted that the monolayer form of these materials are indeed stable,~\cite{Li14p11065,Xingxing14p7071} making them attractive candidates for 2D magnets.  Figure~\ref{fig:Fig1}(a) and (b) show the crystal structure of TMTC monolayers.  The magnetic ions ($A$) form a honeycomb lattice within each layer, and each of them is octahedrally coordinated by six $X$ atoms from its three neighboring ($B_2 X_6$) ligands, with the centers of the hexagons occupied by the $B_2$ groups.

\begin{figure*}
\includegraphics[width=2.0\columnwidth]{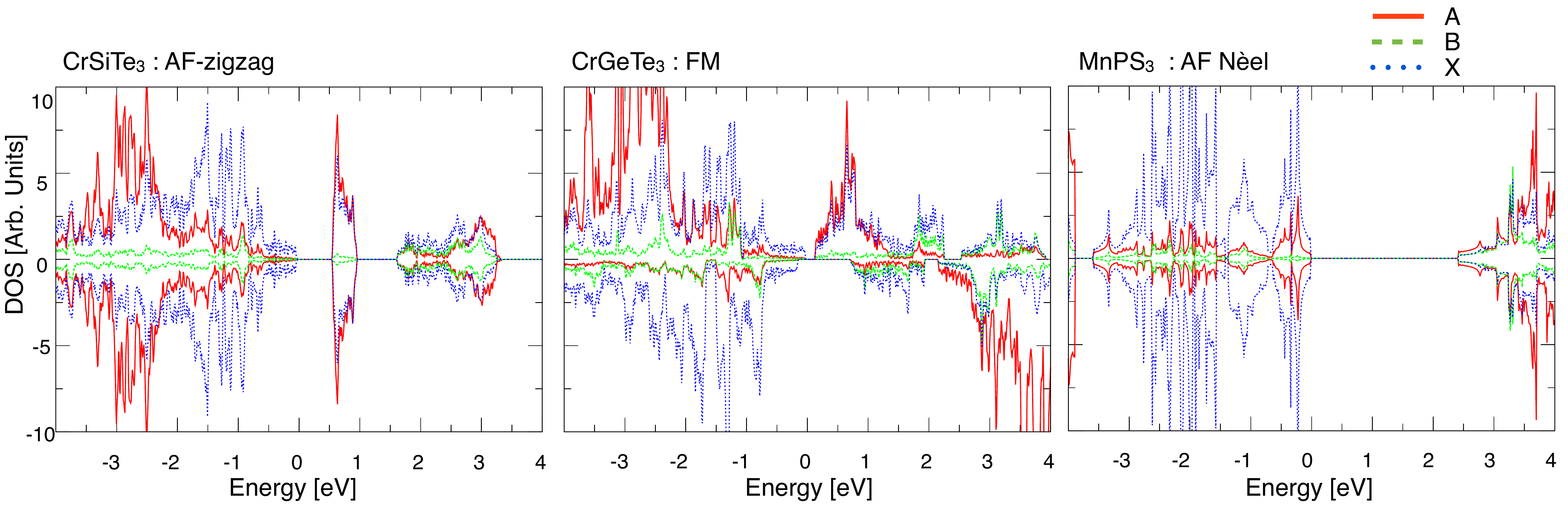}
\caption{\label{fig:FigDOS} (Color online) The partial density of states (PDOS) of the ground states of CrSiTe$_3$ (AF-zigzag), CrGeTe$_3$ (FM) and MnPS$_3$ (AF-N{\'e}el) are shown in (a), (b) and (c) respectively. These are the three unique ground states exhibited by the $ABX_3$ compounds. The PDOS of $A$, $B$ and $X$ are shown using red lines, broken green lines and dotted blue lines respectively. We observe considerable hybridization between the transition metal ($A$) atoms and the chalcogen ($X$) atoms.
}
\end{figure*}

Similar to the crystal structure, the magnetic structure of bulk TMTC also shows 2D characteristics.  It can be understood as FM or AF coupled 2D magnetic layers.  To describe the 2D magnetic structure, we consider the Heisenberg model on a honeycomb lattice,
\begin{equation} \label{H}
H = \sum_{\bracket{ij}}  J_1\vec{S_i} \cdot \vec{S_j} + \sum_{\bracket{\bracket{ij}}}  J_2\vec{S_i} \cdot \vec{S_j} + \sum_{\bracket{\bracket{\bracket{ij}}}}  J_3\vec{S_i} \cdot \vec{S_j} \;.
\end{equation}
where $J_{1,2,3}$ are the exchange interactions between NN, second NN, and third NN spins.  Previous studies have shown that it is necessary to include both $J_2$ and $J_3$ to fit the spin wave dispersion from inelastic neutron scattering data.~\cite{Wildes98p6417,Wildes12p416004}  In addition, considering only $J_1$ would yield either FM or AF-N{\'e}el order, while other magnetic ground states have been found experimentally.  To compute the exchange interactions, we consider the following four possible magnetic ground state: FM, AF-N{\'e}el, AF-zigzag, and AF-stripy, as shown in Fig.~\ref{fig:Fig1} (c)-(f).  
The ground-state phase diagram for our model in Eq.~(1) is shown in Fig.~\ref{fig:J1_vs_J2} as a function of $J_1/J_3$ and $J_2/J_3$.  
Here $J_3$ is assumed to be positive, as it turns out to be the case for all the compounds we studied.
It is clear that not only $J_1$, but also $J_2$ and $J_3$ play an important role in deciding the magnetic ground state.

\section{\label{sec:exchange}Exchange Interactions}

\subsection{Computation details}

With the above observation, the magnetic ground states of $ABX_3$ compounds are examined using DFT employing the projector augmented wave%
\cite{PhysRevB.50.17953,Kresse99p1758,Kresse96p11169} method encoded in Vienna $ab ~ initio$ simulation package~\cite{Kresse96p11169} 
with the generalized gradient approximation in the parameterization  of Perdew, Burke and Enzerhof.~\cite{Perdew96p3865,PhysRevLett.78.1396} 
We use Hubbard $U$ terms (4~eV for Cr and 5~eV for Mn)~\cite{PhysRevB.73.195107, PhysRevB.75.195128} 
to account for strong electronic correlations as suggested by Dudarev {\it et al}.~\cite{PhysRevB.57.1505} 
Our results were qualitatively insensitive to the different $U$'s chosen (2~eV, 4~eV) for the Cr-compounds. For each slab a vacuum region more than 15~{\AA} was used.  
A cutoff energy of 400~eV and a Monkhorst-Pack special $k$-point mesh of 24$\times$14$\times$1 for the Brillouin zone integration was found to be sufficient to obtain the convergence. 
Structural optimizations were performed by fixing the in-plane lattice constants to that of the theoretical bulk lattice constants (see Table I).  
All ions were then relaxed with the relaxation of the electronic degrees of freedom accurate to up to 10$^{-6}$~eV.  

\subsection{Exchange interactions}

For each compound, we optimize the crystal structure for all four spin configurations [see Fig. \ref{fig:Fig1} (c)-(f)] to find the lowest-energy state.  Figure~\ref{fig:FigDOS} shows the partial density of states (DOS) of three representative ground states: AF-zigzag (CrSiTe$_3$), FM (CrGeTe$_3$) and AF-N{\'e}el (MnPS$_3$).  It is evident that there is considerable hybridization between the chalcogen $p$ states and the transition metal $d$ states, further confirming the necessity to include the second and the third NN interaction into consideration.  By integrating the partial DOS in the transition metal atoms for the lowest energy spin configuration we obtain $S_i$ = 2.45 for Mn-compounds and 2.10 for Cr-compounds, with the spins having a variation less than 0.01 between the different spin configurations.

To further extract the $J$'s, we chose to fix the lattice to that of the most energetically favorable spin configuration and computed the energies for different spin configurations. The exchange coupling constants were derived by mapping the DFT energies to the Heisenberg spin Hamiltonian~\eqref{H},
\begin{equation}
\begin{aligned}
E_{\rm FM/N\acute eel} & = E_0+ \left({\pm 3J_1 + 6J_2 \pm 3J_3}\right)|\vec{S}|^{2} ,\\
E_{\rm AF-zigzag/stripy} & = E_0+ \left({\pm J_1 - 2J_2 \mp 3J_3}\right)|\vec{S}|^{2} ,
\label{eq:E}
\end{aligned}
\end{equation}
where $E_0$ is the ground state energy independent of the spin configuration. Using these $J$'s, we also calculated the critical temperature by performing a Monte Carlo simulation of an Ising model on the 2D honeycomb lattice.~\cite{MPX3NS} 

\begin{table}
\caption{\label{tab:table2} Lattice constant $a$, magnetic ground state (GS), exchange coupling constants, and magnetic critical temperature for $ABX_3$ studied.
Critical temperatures are obtained from classical Monte Carlo simulations.
 }
\begin{ruledtabular}
\begin{tabular}{cccccccc}
 & \pbox{20cm}{$a$ \\ (\AA)} & GS &\pbox{20cm}{$J_1$ \\ (meV) }& \pbox{20cm}{$J_2$\\(meV) }& \pbox{20cm}{$J_3$\\(meV) } & \pbox{20cm}{$T_c$\\(K)} 
 \\
\hline
MnPS$_3$ (exp)~\cite{Wildes98p6417}\footnote{ Inelastic neutron scattering was used experimentally to extract the energies.} &5.88& N{\'e}el &0.77 &0.07  & 0.18  &164\
\\
MnPS$_3$&5.88& N{\'e}el & 0.79  & 0.04  & 0.23  &231\
\\
MnPSe$_3$&6.27& N{\'e}el &0.46 &0.03  & 0.19  &147\
\\
MnPSe$_3$ (2$\%$ strain)&6.40& N{\'e}el &0.33 &0.03  & 0.16  &115\
\\
CrSiSe$_3$&6.29& Zigzag & -0.74 &0.0  & 0.43  &92 \
\\
CrSiTe$_3$&6.84& Zigzag &-1.63 &0.08  & 0.71  &160\
\\
CrSiTe$_3$ (1$\%$ strain)&6.91& Zigzag &-1.82 &0.07  & 0.66 &130\
\\
CrSiTe$_3$ (2$\%$ strain)&6.98& Zigzag &-1.99 & 0.07 & 0.60 &72\
\\
CrSiTe$_3$ (3$\%$ strain)&7.04& FM &-2.16 &0.05  & 0.54 &111\
\\
CrSiTe$_3$ (4$\%$ strain)&7.11& FM &-2.29 &0.05  & 0.50  &158\
\\
CrGeTe$_3$&6.91& FM &-1.88 &0.20  & 0.22  &106\
\\
\end{tabular}
\end{ruledtabular}
\end{table}

\begin{figure*}
\includegraphics[width=2.0\columnwidth]{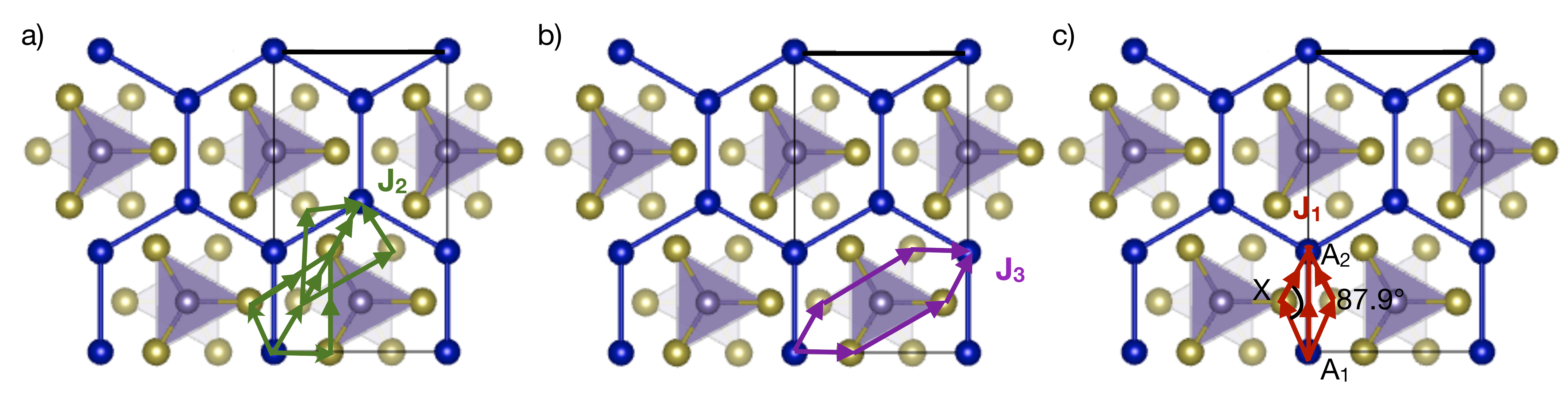}
\caption{\label{fig:Fig3} (Color online) Top view of monolayer of $ABX_3$. Blue, violet and yellow represents $A$, $B$ and $X$ ions respectively. The five possible paths for second NN interaction is show in (a) and the third NN interaction is shown in (b). The two different NN hopping paths between two transition metal atoms at different sites 
($A_1$ and $A_2$) are shown in (c) with the direct exchange as a hopping between the transition metal orbitals and the superexchange interaction characterized by hopping between the transition metals through the $X$ atom. 
}
\end{figure*}

The magnetic ground state, the computed $J$'s, along with the critical temperature for each compound are listed in Table I. The locations of the ground state of all the compounds studied are labeled in the phase diagram in Fig.~\ref{fig:J1_vs_J2}. We see that both MnPS$_3$ and MnPSe$_3$ are deep inside the AF-N{\'e}el phase. The calculated values of $J_{1,2,3}$ for monolayer MnPS$_3$ agree excellently with the experimental data for the bulk system,~\cite{Wildes98p6417} which validates our calculation. We also find that both CrSiTe$_3$ and CrSiSe$_3$ are in the AF-zigzag phase with the former lying close to the boundary of  AF-zigzag and FM phase.  This is different from the FM ground state reported for bulk CrSiTe$_3$,~\cite{Carteaux95p251,Casto15p} which will be addressed later.  Finally, of all the compounds studied, CrGeTe$_3$ is the only one that has a FM ground state in its unstrained monolayer form.

\begin{figure}
\includegraphics[width=\columnwidth]{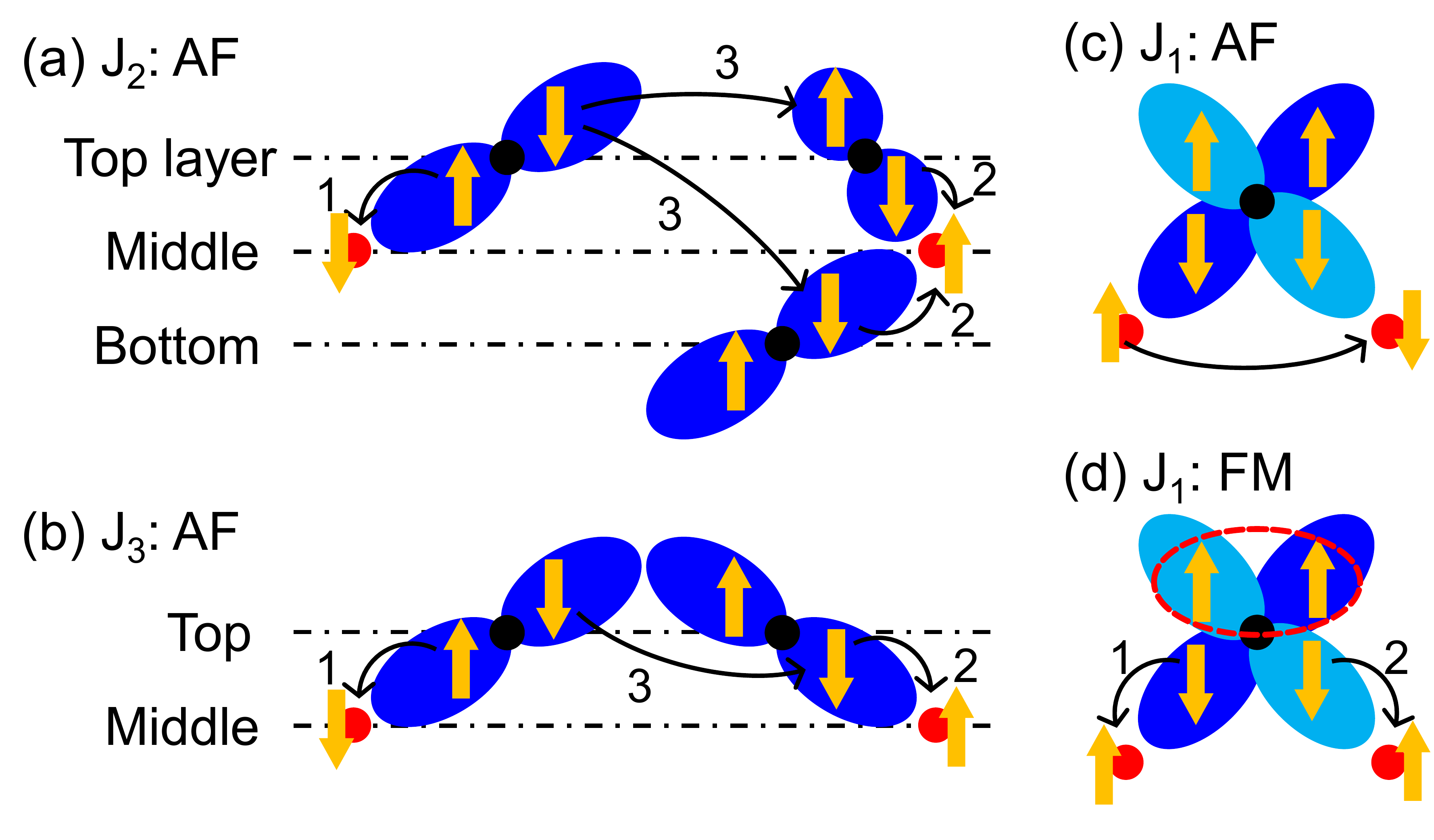}
\caption{\label{fig:Js} (Color online) Main contributions of virtual electron excitations to magnetic interactions.
Virtual electron excitations from two $X$ anions to a pair of (a) second NN and (b) third NN $A$ ions. 
(c) Direct excitations between neighboring TM ions (red dots), resulting in AF $J_1$.
(d) Excitations from two orthogonal orbitals on a $X$ anion (black dots) to a neighboring pair of $A$ ions.
Because of the Hund coupling acting in excited states as indicated by a broken circle, this process gives rise to FM $J_1$.
The actual sign of exchange interactions results from the competition between FM and AF contributions for $J_1$. 
Numbers indicate the typical order of the first half perturbation processes for each contribution. 
}
\end{figure}

We note that $J_3$ is significantly large. This corroborates the decision to include more than just the NN interaction. Ignoring it (and $J_2$) had previously yielded a different ground state (FM) for CrSiTe$_3$ monolayers in previous studies.~\cite{Xingxing14p7071,Chen15p60} In our calculation for monolayer CrSiTe$_3$, FM is indeed lower in energy than AF-N{\'e}el. The energy difference between the two magnetic states was found to be similar to what was reported in Li {\it et al.}~\cite{Xingxing14p7071}, when the same $U$ was chosen. But crucially, we find that AF-zigzag is even lower in energy than FM, and hence is the magnetic ground state.

Now that we have shown that the interactions have to be included up to the third NN not only to interpret the neutron diffraction data, 
but also to get the correct magnetic ground state, the imminent task is to understand the microscopic origin of the different $J$'s. 
We first note that  $J_2$ and $J_3$ are always AF. Furthermore, the value of $J_2$ is found to be smaller than $J_3$. 
Both of these findings are consistent with previous reports on MnPS$_3$ and its Fe derivative, FePS$_3$.~\cite{Wildes98p6417,Wildes12p416004} 
These observations can be understood by analyzing  the crystal structure. 
Figure~\ref{fig:Fig3}  shows the possible hybridization paths connecting $A$ site ions. 
For the second NN and the third NN $A$ site pairs, electrons hop through two $X$ anions [Fig.~\ref{fig:Fig3} (a) and (b)]. 
For this reason, $J_2$ and $J_3$ might be regarded as super-superexchange interactions.
Based on the geometry and the $X$ anion $p$ states, we expect $J_2$ to be weakly AF because it involves small $X$-$X$ hybridizations [Fig.~\ref{fig:Js} (a)]. On the other hand, $J_3$ involves two $X$ anions on the same plane, either top layer or bottom layer. Hence, there is strong hybridization of the $p$ states [Fig.~\ref{fig:Js} (b)], resulting in a strongly AF $J_3$.

The NN exchange $J_1$, on the other hand, shows a large variation from compound to compound. 
The variation is so large that it even changes the sign going from the Mn compounds to the Cr compounds (see Table I). 
This behavior comes from a unique crystal structure, which naturally gives rise to two competing interactions, i.e., the direct exchange and superexchange. 
The direct exchange originates from direct electron hopping between the NN $A$ sites [see Figs. \ref{fig:Fig3} (c) and  \ref{fig:Js} (c)]. 
For the Mn compounds, this exchange is robustly AF as the Mn ions are in the half-filled high-spin $d^5$ state. 
For the Cr compounds, the AF direct exchange is weakened by a FM component as Cr ions have partially-filled $d$ shell.~\cite{PhysRevB.87.014418} 
The superexchange interaction is mediated through the $X$ ions [see Fig. \ref{fig:Fig3} (c)]. 
As the $A_1$-$X$-$A_2$ angle is close to 90$^\circ$, this interaction is FM.~\cite{Goodenough55p564,Goodenough58p287}  
It is important to note that for the superexchange interaction 
two electrons must excite from $X$ anion $p$ states to neighboring $A$ $d$ states [see Fig. \ref{fig:Js} (d)]. 
Since the electron excitation energy is large for the Mn compounds [Fig.\ref{fig:FigDOS}(c)], reflecting closed $d$ shell on Mn ions, 
the superexchange is expected to play a minor role compared with the direct exchange. 
On the other hand for the Cr compounds, the superexchange could play a dominant role. 

To confirm the distinct role of the superexchange mechanism's contribution to $J_1$, 
we examine the magnetization of a chalcogen ion between two ferromagnetically coupled transition metal ions  
as its magnitude is a good indication of the strength of the superexchange interaction.~\cite{Mazurenko07p224408} 
Our DFT calculation for the Cr compounds in the FM metastable state showed the total magnetization of the chalcogen ions is $\sim 0.6$ $\mu_B$ per unit cell and hence significant. 
On the other hand, for the Mn compounds, the magnetization of the chalcogen ions are an order of magnitude smaller, which is consistent with our finding of an AF $J_1$, 
and the presence of a large electron excitation energy from $X$ $p$ to Mn $d$.
The net result is that $J_1$ becomes AF for Mn compounds because of the dominance of AF direct exchange over the FM superexchange, 
while the FM superexchange wins over the AF direct exchange making $J_1$ strongly FM for Cr compounds (see Table I).
Hence, depending on the transition metal ions involved, a significant competition is expected between FM and AF components, which could lead to a plethora of magnetic states.

This competition is further verified by applying a uniform tensile strain on MnPSe$_3$ and CrSiTe$_3$. 
It is important to note that both direct exchange and superexchange do not change sign as we strain the system, 
but the former decreases more rapidly than the latter as the atomic distances increase by a tensile strain. 
As a consequence, $|J_1|$ for MnPSe$_3$ decreases with strain,  where as $|J_1|$ for CrSiTe$_3$ increases with strain (see Table I). 
This confirmed the presence of competing exchange interactions.


This result immediately suggests the possibility of tuning the magnetic ground state using strain.
Here, we consider monolayer CrSiTe$_3$ as our prototype system and use strain as a knob to change the different $J$'s. 
CrSiTe$_3$ is an ideal candidate for this study as it lies close to the FM and AF-zigzag phase boundary. 
With a tensile strain, $J_2$ and $J_3$ are both expected to decrease in magnitude as the atomic distances are increased. 
While the effect of strain on $J_1$ is subtler, it is expected to increase in magnitude for small strains. 
Not surprisingly, an application of $\sim 3 \%$ strain leads to a magnetic phase transition with ferromagnetism becomes the magnetic ground state (see Fig.~\ref{fig:J1_vs_J2}).
Strain also has a direct impact on the critical temperatures. 
Once the FM ground state is realized, the critical temperature $T_c$ can be further enhanced with strain. 
As shown in Table I, $T_c$ goes up to 158~K for $\sim 4 \%$ strain. 
With this strong dependence of critical temperature on the applied strain, it might be even possible to engineer room temperature ferromagnetic behavior, for large values of strain.~\cite{Chen15p60}


\section{\label{sec:discussion}Bulk magnetic order}

So far we have only considered the magnetic properties of monolayer TMTCs.  
One of the important finding is that monolayer of CrSiTe$_3$ has an AF-zigzag ground state, whereas in bulk it is reported to be FM from neutron scattering experiments.~\cite{Carteaux95p251,Casto15p}  
To understand this change in magnetic structure when we go from monolayer to bulk, we calculated the magnetic ground state of bulk CrSiTe$_3$. 
We find that for the experimental out-of-plane lattice constant (21.0 {\AA}), the intralayer magnetic ordering is FM in nature and not AF-zigzag. 
This switching of the intralayer magnetic ordering from AF-zigzag to FM as we go from monolayer to bulk is very sensitive to the out-of-plane lattice constant of the bulk system. 
If we increase the out-of-plane lattice constant to 22.8 {\AA}, the intralayer coupling prefers AF-zigzag. 
This is because bulk CrSiTe$_3$ has $ABC$ stacking, thus an in-plane AF-zigzag spin configuration costs more energy compared to a FM spin configuration 
when the interlayer exchange interaction becomes strong. 

However, we also find that in bulk, AF interlayer coupling is preferred over FM interlayer coupling by 10.6~meV per Cr atom. 
Analyzing the interlayer interactions between the Cr atoms, it is dominated by super-superexchange. 
Based on the geometry and because of the presence of large chalcogen atoms, we expect this interaction to be comparable to $J_3$ and hence significant. 
As the mechanism for interlayer super-superexchange coupling is similar to that of the intralayer super-superexchange coupling previously discussed for $J_2$ and $J_3$, 
it is not surprising that these interactions are AF in nature. 
But this contradicts experimental findings of bulk ferromagnetism in CrSiTe$_3$.~\cite{Carteaux95p251,Casto15p}  

One possible source for this discrepancy is the absence of the dipole-dipole interaction in DFT calculations, as discussed previously. ~\cite{Koo09p9051,Kabbour12p11915}
By introducing the spin-orbit coupling, we have confirmed that there is an out-of-plane easy axis anisotropy, in accordance with experiments.~\cite{Carteaux95p251,Casto15p} 
With this easy axis, the interlayer FM arrangement can become energetically more favorable than the AF arrangement, because of the dipole-dipole interaction. ~\cite{Kabbour12p11915} We also note that for MnPS$_3$ the dipole-dipole interaction is negligible due to the AF ordering within each plane. This could explain why our DFT results show excellent agreement with the bulk experimental results for the Mn compounds. Nonetheless, the actual mechanism for ferromagnetism in bulk CrSiTe$_3$ remains an open question. 

\section{\label{sec:summary}Summary}

In this paper, we studied the magnetic properties of monolayers of van der Waals transition-metal trichalcogenides $ABX_3$ using density functional theory. 
In order to understand the rich magnetic behavior observed in these systems, 
we derived local spin models using the DFT energy of the magnetic ground state and metastable excited states. 
Because of the extended nature of the $p$ state of the chalcogen atoms, second nearest-neighbor and third nearest-neighbor interactions are found to play significant roles.  Specifically, we find that monolayer CrSiTe$_3$ is an antiferromagnet with a zigzag spin texture due to significant contribution from $J_3$, 
whereas CrGeTe$_3$ is a ferromagnet with a Curie temperature of 106 K.  
Detailed analyses on the magnetic interactions led us to predict that 
monolayers CrSiTe$_3$ can be made ferromagnetic with the application of a moderate uniform in-plane tensile strain of 3$\%$, which is experimentally feasible. 
Our studies demonstrate transition-metal trichalcogenides $ABX_3$ are possible candidates for spintronic applications; 
especially CrGeTe$_3$ and strained CrSiTe$_3$ are promising for two-dimensional ferromagnetic semiconductors.
The magnetic ordering of bulk CrSiTe$_3$, however, remains an open question.


\section*{Acknowledgments}

We are grateful to David Mandrus and Jiaqiang Yan for bringing CrSiTe$_3$ to our attention, and to Kai Xiao, Zheng Gai, and Travis J. Williams for sharing their experimental data before publication. We would also like to thank Wenguang Zhu, Ji Feng and Xiao Li for their computational input. This work is supported by the Air Force Office of Scientific Research under Grant No.~FA9550-12-1-0479 and FA9550-14-1-0277, and by the National Science Foundation under Grant No.~EFRI-1433496.  S.O. acknowledges support by the U.S. Department of Energy, Office of Science, Basic Energy Sciences, Materials Sciences and Engineering Division.


\begin{thebibliography}{45}
\bibitem [{MPX()}]{MPX3NS}%
  \BibitemOpen
  \href@noop {} {\bibinfo  {journal} {Using $J_1$, $J_2$, and $J_3$ hopping
  energies determined from DFT, we conducted a Monte Carlo simulation of an
  Ising model on the 2D honeycomb lattice with periodic boundary conditions.
  The calculations proceeded according to the standard Metropolis algorithm.
  The system was annealed from a uniformly random spin distribution at 400K
  down to absolute zero in half Kelvin steps. Size effects were accounted for
  by repeating trials on systems from 16 atomic sites up to 256 atomic sites.
  Textbook histogram methods were then used to determine the critical
  temperature of the various parameterizations listed in Table I.}}\BibitemShut
  {Stop}%
\end{thebibliography}

%

\end{document}